# Macroscopic quantum phenomena and quantum computing

Jian-Qiang You[1,2,*]

[1]*Zhejiang Key Laboratory of Micro-Nano Quantum Chips and Quantum Control, School of Physics, and State Key Laboratory for Extreme Photonics and Instrumentation, Zhejiang University, Hangzhou 310027, China*

[2]*College of Optical Science and Engineering, Zhejiang University, Hangzhou 310027, China*

[*]*Email: jqyou@zju.edu.cn*

This News&Views article provides a perspective on the 2025 Nobel Prize in Physics, including the groundbreaking discovery of macroscopic quantum tunneling and energy quantization in superconducting circuits, the history and causes giving rise to this breakthrough, and its impact on subsequent progress in quantum computing.

A bulk superconductor contains a vast number of paired electrons, known as Cooper pairs, which collectively condense into a single quantum state. This state is described by the wavefunction $\sqrt{\rho}e^{i\phi}$, where $\rho$ represents the density of Cooper pairs. In the absence of an applied current or magnetic field, the phase $\phi$ remains uniform across the superconductor. When two bulk superconductors are separated by a thin insulating layer, a supercurrent flows through the junction, expressed as $I = I_c \sin\varphi$. Here $I_c$ is the critical current that the junction can sustain, and $\varphi = \phi_L - \phi_R$ is the phase difference across the junction (see Fig. 1a). This phenomenon, commonly referred to as the Josephson effect, was first predicted by Brian Josephson and later experimentally validated [1], earning him the 1973 Nobel Prize in Physics for this groundbreaking discovery.

One of the most significant applications of the Josephson effect is the highly sensitive detection of weak magnetic fields using a superconducting quantum interference device (SQUID). A SQUID consists of a superconducting loop interrupted by two Josephson junctions and is influenced by a magnetic field threading the loop. Another key application is in establishing a precise voltage standard, achieved by utilizing the voltage difference across the junction, which is directly related to the time variation of the phase drop across the junction.

In early development, Josephson junctions were constructed with relatively large dimensions, allowing a significant number of Cooper pairs to tunnel through. As a result, superconducting circuits based on these junctions exhibited classical behavior. However, in the 1980s, researchers realized that both reducing the size of the junction and improving its quality could suppress environmental influences, enabling the junction to exhibit quantum mechanical behavior. This insight led to the discovery of macroscopic quantum mechanical tunneling and energy quantization in superconducting circuits [2,3]. For this groundbreaking achievement, John Clarke of UC Berkeley, Michel Devoret of Yale University, and John Martinis of UC Santa Barbara were jointly awarded Nobel Prize in Physics this year.

The supercurrent through a Josephson junction is associated with a coupling energy $U(\varphi) = E_J\,(1-\cos\varphi)$, where $E_J$ is the Josephson energy, determined by $I_c = (2e/\hbar)E_J$. A Josephson junction can also be regarded as a capacitor with capacitance $C_J$, meaning that a Cooper pair tunneling through the junction must overcome a charging energy $E_c = (2e)^2/2C_J$. Including this charging energy, the Hamiltonian of the Josephson junction is given by $H = E_c N^2 + E_J\,(1-\cos\varphi)$, where $N$ represents the number of Cooper pairs tunneling through the junction. This Hamiltonian provides a clear framework for understanding both the classical and quantum behaviors of the junction. For a large Josephson junction, the capacitance $C_J$ is significant, making the charging energy $E_c$ negligible. In this case, the Hamiltonian simplifies to the Josephson coupling energy $E_J(1-\cos\varphi)$, and the system behaves classically, as was typical in early Josephson junctions. However, when the junction size is reduced, as in [2,3], the charging energy becomes significant. With a larger $E_c$, fewer Cooper pairs can tunnel through the junction, transforming it into a quantum junction. Here, the conjugate variables $N$ and $\varphi$ obey the canonical commutation relation $[\varphi, N] = i$. Consequently, the Josephson junction behaves like a quantum "particle" with an effective mass $M \equiv C_J(\hbar/2e)^2$ in a periodic potential. The tunneling of Cooper pairs through this quantum junction differs fundamentally from that in a conventional Josephson junction [2], as it primarily involves quantum tunneling between distinct macroscopic quantum states—a phenomenon known as the macroscopic quantum tunneling.

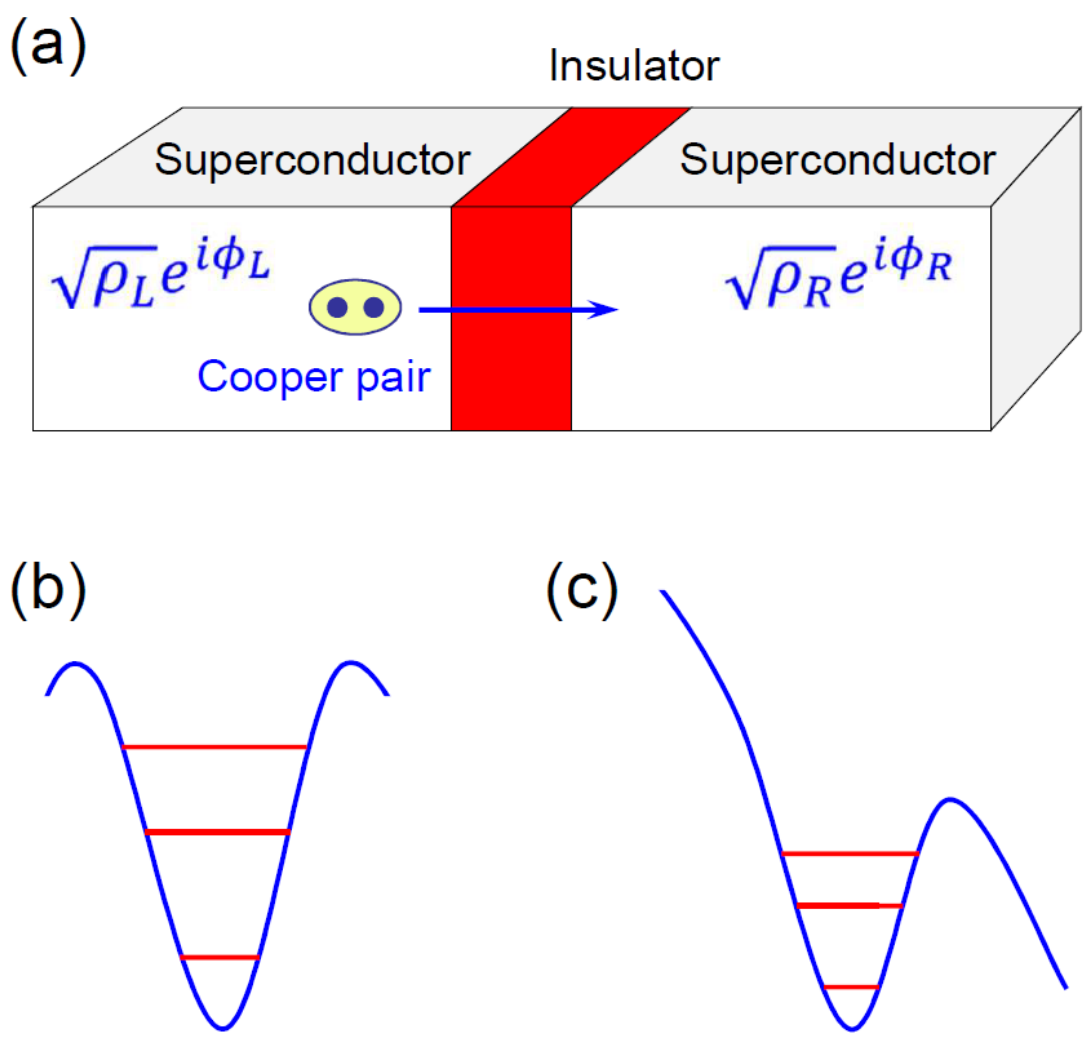

**Fig. 1** (a) A Josephson junction consisting of two bulk superconductors separated by a thin insulating layer. (b) Discrete energy levels in a given well of the periodic potential related to Josephson coupling energy. More energy levels can form in the well by reducing the charging energy of the junction. (c) Discrete energy levels in the well when the Josephson junction is biased by a current.

This macroscopic quantum system exhibits energy quantization, meaning discrete energy levels form within wells of the periodic potential (Fig. 1b). To observe this quantization, experiments in [3] employed a current-biased Josephson junction. The bias current introduces an additional potential proportional to $\varphi$, giving rise to a tilted washboard potential for the junction. The macroscopic quantum state in the highest energy level of a given well faces a low barrier (Fig. 1c), enabling faster tunneling compared to lower energy states. By irradiating the junction with a microwave field of appropriate frequency and tuning the bias current to adjust the potential profile, tunneling events from specific excited states, such as the first and second excited states, can be selectively measured. Analyzing these tunneling characteristics using the corresponding theoretical model allows for the identification of discrete energy levels within the well [3].

The number of energy levels in a well depends on the competition between the charging energy $E_c$ and the Josephson energy $E_J$. Increasing the ratio $E_c/E_J$ by reducing the junction size increases the spacing between energy levels, thereby decreasing their number in the well. In summary, the experiments in [2,3] demonstrated the quantum behavior of a macroscopic variable in a current-biased Josephson junction by revealing its energy quantization and macroscopic quantum mechanical tunneling.

The experiments in [2,3] had a profound impact, catalyzing the development of quantum information processing using superconducting circuits. The first major breakthrough came with the implementation of a charge qubit, achieved by demonstrating quantum oscillations—coherent superpositions of macroscopic quantum states—in a superconducting circuit known as the Cooper-pair box [4], despite its short decoherence time of approximately 10 ns. Subsequently, quantum oscillations were observed in a phase qubit [5], which utilized a current-biased Josephson junction similar to those in [2,3] but with improved design and junction quality. The decoherence time was estimated to be about 5 μs for this phase qubit, significantly longer than that of the charge qubit in [4]. These findings were followed by the observation of quantum oscillations in other types of superconducting qubits, such as the flux qubit, though their decoherence times remained insufficient for implementing extensive quantum information processing operations.

A pivotal moment arrived in 2007 when researchers discovered that charge noise dominated in flux qubits. To mitigate this, a large capacitor was shunted to the small junction in the flux qubit, reducing the sensitivity of the resulting C-shunt flux qubit to charge noise [6]. This approach was later adapted for charge qubits, leading to the development of the C-shunt charge qubit, or transmon [7]. Further refinements enabled the transmon to be conveniently coupled to adjacent qubits [8], resulting in the modified transmon, or Xmon, which became widely adopted in multi-qubit quantum processors due to its simplicity and scalability. These advancements culminated in landmark experiments [9,10] that sampled the output distribution of random quantum circuits using over fifty superconducting qubits, demonstrating quantum advantage over the most advanced classical computers. This marked a significant milestone in the field of quantum information processing.

The shunt capacitor introduces an innovative approach to enhancing the quantum coherence of superconducting circuits. Thanks to this advancement, superconducting qubits can now achieve decoherence times as long as approximately 1 millisecond, which is five orders of magnitude longer than the initial charge qubit in [4]. For quantum computing, the ultimate objective is to construct a universal fault-tolerant quantum machine comprising a large number of logical qubits equipped with sophisticated error-correction capabilities. Achieving this goal necessitates further improvements in the quantum coherence of superconducting qubits, as a single logical qubit is typically encoded using multiple physical qubits (e.g., in surface codes). Additionally, implementing quantum algorithms with practicality demands even higher levels of quantum coherence in superconducting qubits. For example, the renowned Shor algorithm for integer factorization includes a critical component known as the quantum Fourier transformation. While the original algorithm relies on controlled-phase gates between every pair of qubits in a multi-qubit processor, practical large-scale processors only support nearest-neighbor interactions between qubits. Consequently, the algorithm must be adapted to function within a quantum processor limited to nearest-neighbor interactions, significantly increasing the number of required one-qubit and two-qubit operations.

To further improve the quantum coherence of superconducting qubits, it is essential to identify the dominant remaining noise in the circuits after the impact of charge noise has been significantly mitigated through the use of shunt capacitance. From a fundamental perspective, understanding the mechanisms behind each type of quantum noise in superconducting circuits is crucial. Scientifically, this is a challenging endeavor that requires substantial effort, as unraveling the mechanisms of quantum noise has long been a formidable and persistent problem in physics.

However, solving this problem would pave the way for the realization of a large-scale superconducting quantum machine capable of implementing universal fault-tolerant quantum computing.

**Acknowledgments**

This work was supported by the National Natural Science Foundation of China (92265202) and the National Key R&D Program of China (2022YFA1405200).